\documentstyle[12pt]{article}
\newcommand{\la}{\label}

\newcommand{\F}{{\cal F}}
\newcommand{\D}{{\cal D}}

\newcommand{\be}{\begin{equation}}
\newcommand{\ee}{\end{equation}}
\newcommand{\ba}{\begin{eqnarray}}
\newcommand{\ea}{\end{eqnarray}}
\newcommand{\bastar}{\begin{eqnarray*}}
\newcommand{\eastar}{\end{eqnarray*}}
\newcommand{\half}{{1 \over 2}}
\begin{document}
\begin{titlepage}

\vskip 0.4truecm
 
\begin{center}
{ 
\bf \large \bf PARTIAL DUALITY IN SU(N) \\ \vskip 0.3cm
                 YANG-MILLS THEORY \\
}
\end{center}
 
\vskip 1.0cm
 
\begin{center}
{\bf Ludvig Faddeev$^{* \sharp}$ } {\bf \ and \ } {\bf 
Antti J. Niemi$^{** \sharp}$ } \\
\vskip 0.3cm

{\it $^*$St.Petersburg Branch of Steklov Mathematical
Institute \\
Russian Academy  of Sciences, Fontanka 27 , St.Petersburg, 
Russia$^{\ddagger}$ } \\

\vskip 0.3cm

{\it $^{**}$Department of Theoretical Physics,
Uppsala University \\
P.O. Box 803, S-75108, Uppsala, Sweden$^{\ddagger}$ } \\

\vskip 0.3cm

{\it $^{\sharp}$Helsinki Institute of Physics \\
P.O. Box 9, FIN-00014 University of Helsinki, Finland} \\

\vskip 0.3cm

{\it $^{**}$ The Mittag-Leffler Institute  \\
Aurav\"agen 17, S-182 62 Djursholm, Sweden}

\end{center}

\vskip 0.5cm
\rm
Recently we have proposed a set of variables
for describing the infrared limit of four dimensional 
$SU(2)$ Yang-Mills theory. Here we extend 
these variables to the general case of four 
dimensional $SU(N)$ Yang-Mills theory. 
We find that the $SU(N)$ connection $A_\mu$ decomposes 
according to irreducible representations 
of $SO(N-1)$, and the curvature two form $F_{\mu\nu}$ 
is related to the symplectic Kirillov two forms that 
characterize irreducible representations 
of $SU(N)$. We propose a general class of nonlinear
chiral models that may describe stable, 
soliton-like configurations with nontrivial 
topological numbers.

\noindent\vfill
 
\begin{flushleft}
\rule{5.1 in}{.007 in} \\
$^{\ddagger}$  \small permanent address \\ \vskip 0.2cm
$^{*}$ \small supported by Russian Fund for Fundamental Science
\\ \vskip 0.2cm 
$^{**}$ \small Supported by NFR Grant F-AA/FU 06821-308
\\ \vskip 0.3cm
$^{*}$ \hskip 0.2cm {\small  E-mail: \scriptsize
\bf FADDEEV@PDMI.RAS.RU and FADDEEV@ROCK.HELSINKI.FI } \\
$^{**}$  {\small E-mail: \scriptsize
\bf NIEMI@TEORFYS.UU.SE}  \\
\end{flushleft}
\end{titlepage}

Recently \cite{fad1} we have proposed a novel  
decomposition of the four dimensional $SU(2)$ 
Yang-Mills connection $A^a_\mu$. In addition of a 
three component unit vector $n^a$, it involves 
an abelian gauge field $C_\mu$ and a complex scalar 
$\phi = \rho + i \sigma$. The fields $C_\mu$ and $\phi$
determine a $U(1)$ multiplet under $SU(2)$ gauge 
transformations in the direction $n^a$,
\be
A^a_\mu \ = \ C_\mu n^a \ + \ \epsilon_{abc} 
\partial_\mu n^b n^c \ + \ \rho  \partial_\mu n^a \ + \ 
\sigma \epsilon^{abc} \partial_\mu n^b n^c
\la{su2}
\ee
In four dimensions this decomposition is complete 
in the sense that it reproduces the Yang-Mills equations 
of motion \cite{fad1}. This is already suggested 
by the number of independent fields: 
if we account for the $U(1)$ invariance, 
(\ref{su2}) describes $D+2$ field degrees of freedom.
For $D=4$ this equals $3(D-2)$, the number 
of transverse polarization degrees of freedom described 
by a $SU(2)$ connection in $D$ dimensions.
Furthermore, if we properly specify the component  
fields, (\ref{su2}) reduces to several 
known $SU(2)$ field configurations.
As an example, if we set $A_0 = C_\mu = 
\rho = \sigma = 0$ and specify $n^a$ to 
coincide with the (singular) radial
vector $x^a/r$, the parametrization (\ref{su2}) yields  
the (singular) Wu-Yang monopole configuration \cite{wu}. 
As another example, if we specify a rotation symmetric 
configuration with $C_i = C x^i$, and $C_0$, $C$, $\phi$ 
to depend on $r$ and $t$ only, and set $n^a = x^a / r$, 
(\ref{su2}) reduces to Witten's Ansatz for 
instantons \cite{witten}.

Here we wish to generalize (\ref{su2}) to $SU(N)$ We shall  
argue that exactly in four dimensions the
$SU(N)$ Yang-Mills connection admits the following 
decomposition  
\be
A^a_\mu \ = \ C^i_\mu m^a_i \ + \ f^{abc} \partial_\mu m^b_i
m^c_i \ + \ \rho^{ij} \ f^{abc} \partial_\mu m^b_i m^c_j \ + \
\sigma^{ij} \ d^{abc} \partial_\mu m^b_i m^c_j
\la{A1}
\ee
With $i=1,...,N-1$ we label the Cartan subalgebra.
We shall construct the $N-1$ mutually orthogonal 
unit vector fields $m^a_i$ (with $a=1,...,N^2-1 = 
Dim[SU(n)]$ in the following) so that they describe 
$N(N-1)$ independent variables. The combination
\be
A^a_\mu \ = \ C^i_\mu m^a_i \ + \ f^{abc} 
\partial_\mu m^b_i
m^c_i 
\la{cho}
\ee
is the $SU(N)$ Cho connection \cite{cho}, under the $N-1$ 
independent gauge transformations generated by the 
Lie-algebra elements 
\be
\alpha_i \ = \ \alpha_i m^a_i T^a
\la{U1}
\ee
the vector fields $C^i_\mu$ transform as U(1) connections 
\[
C^i_\mu \ \to \ C^i_\mu + \partial_\mu \alpha_i
\]
Consequently these fields describe $(D-2)(N^2-1)$ 
independent variables. The 
$\phi_{ij} = \rho_{ij} + i \sigma_{ij}$ are $N(N - 1)$ 
independent complex scalars, mapped onto each other by 
the $U(1)$ gauge transformations (\ref{U1}). As a consequence 
$(C^i_\mu , \phi_{ij})$ can be viewed as a collection of abelian 
Higgs multiplets. We shall find that the fields $\rho_{ij}$ and
$\sigma_{ij}$ decompose according to the traceless symmetric 
tensor, the antisymmetric tensor, the vector and the singlet 
representations of $SO(N-1)$. In $D$ dimensions 
(\ref{A1}) then describes 
\[
N(N-1) \ + \ (D-2) \dot (N-1) \ + \ N(N-1) \ = \ 
2 N^2 \ + \ (D-4) N \ + \ (2-D)
\]
independent variables. In {\it exactly $D=4$} this 
coincides with $(D-2) (N^2 -1)$ which is the number of 
independent transverse variables described by a $SU(N)$
Yang-Mills connection $A^a_\mu$.  

We also observe that for $D=3$, the number of independent variables 
in (\ref{A1}) coincides with the dimension of the $SU(N)$
gauge orbit, {\it independently} of the Yang-Mills
equations of motion.

We note that a decomposition of the $SU(N)$ connection has 
been recently considered by Periwal \cite{periwal}. 
It appears that his results are different from ours.

We proceed to the justification of the decomposition
(\ref{A1}). For this we need a number of group theoretical 
relations. Some of these relations seem to be novel, 
suggesting that further investigations of  (\ref{A1}) 
could reveal hitherto unknown structures.

The defining representation of the $SU(N)$ 
Lie algebra consists of $N^2 - 1 $ traceless 
hermitean $N \times N$ matrices $T^a$ with 
\[
T^a T^b \ = \ \frac{1}{2N} \delta^{ab} \ + \ \frac{i}{2} 
f^{abc} T^c 
+ \half d^{abc} T^c
\]
normalized to
\[
(T^a, T^b) \ \equiv \ Tr( T^a T^b ) \ = \ \half \delta^{ab}
\]
The $f^{abc}$ are (completely antisymmetric and real) 
structure constants. The $d^{abc}$ are the 
(completely symmetric) coefficients
\[
\half d^{abc} \ = \ Tr(T^a \{T^b , T^c \}) \ 
\equiv \ Tr(T^a ( T^b T^c + T^c T^b ))
\]
and we note that in the defining representation 
we can select the Cartan subalgebra so that 
\be
\sum_{i,j=1}^{N-1} d^{ijk} d^{ijl} \ = \ \frac{ 2 (N-2) }{
N } \delta_{kl}
\la{dijk}
\ee

For any four matrices $A, \ B, \ C, \ D$ 
we have
\[
Tr ( \ [A,B] \{ C,D \} + [A,C] \{ B,D \} + 
[A,D] \{ B,C \} \ )
\ = \ 0
\]
and
\[
Tr( \ [ A,B ][C,D] + \{ A,C \} \{ B,D \} - 
\{ A,D \} \{ B,C \} \ ) \ = \ 0
\]
Hence, if we introduce the matrices 
$(\F^a)^{bc} = f^{abc} $ (which define the  
adjoint representation) and $(\D^a)^{bc} 
= d^{abc}$, we have
\be
[ \F^a , \D^b ] \ = \ - f^{abc} \D^c
\la{fd}
\ee
and
\be
[ \D^a , \D^b ]_{cd} \ = \ f^{abe} \F^e_{cd} \ + \ \frac{2}{N}
(\delta^{ad} \delta^{bc} \ - \ \delta^{ac} \delta^{bd} )
\la{dd}
\ee
Note that the $\D^a$ are traceless.

We conjugate the Cartan matrices $H_i$ of the defining
representation by a generic element $g \in SU(N)$.
This produces set of Lie algebra valued vectors 
\be
m_i \ = \ m^a_i T^a \ = \ g H_i g^{-1}
\la{m}
\ee
an over-determined set of coordinates on the orbit
$SU(N)/U(1)^{N-1}$: By construction the $m_i^a$
depend on $N(N-1)$ independent variables, they are
orthonormal, 
\[
(m_i, m_j) \ = \ m^a_i m^a_j \ = \ \delta_{ij}
\]
and it is straightforward to verify that
\ba
\left[ m_i , m_j \right] \ & = & \ 0 \\
\{ m_i , m_j \} \ & = & \ d^{ijk} m_k 
\la{redu}
\\
Tr ( m_i \partial_\mu m_j ) \ & = & \ 
(m_i , \partial_\mu m_j ) \  =  \ 0
\ea

We can also show that  
\be
- f^{acd} m^c_i f^{deb} m^e_i \ = \ \delta^{ab} - m^a_i m^b_i 
\la{proj1}
\ee
is a projection operator. This can be 
verified {\it e.g.} by using explicitly the 
defining representation of $SU(N)$.
The result follows since the weights of $SU(N)$ have 
the same length and the angles between different weights are
equal. Notice that (\ref{proj1}) can also be represented 
covariantly, using the $SU(N)$ permutation operator
\be
- [ m_i , T^d ] \otimes [m_i , T^d ] \ = \ 
T^b \otimes T^a \ - \ m_i \otimes m_i 
\la{proj2}
\ee

We now consider the Maurer-Cartan one-form
\[
dg g^{-1} \ = \ \omega_{a\mu} T^a dx^\mu
\]
We find 
\be
dm_i \ = \ [ \omega , m_i ]
\la{m2}
\ee
We now recall that each Cartan matrix $H_i$ can 
be used to construct a symplectic (Kirillov) two-form on 
the orbit $SU(N) / U(1)^{N-1}$,
\be
\Omega^i \ = \ Tr ( \ H_i [ g^{-1} d g , g^{-1} d g ] \ ) 
\la{kirillov}
\ee
\[
d\Omega^i \ = \ 0
\]
The $\Omega^i$ are related to the representations of $SU(N)$;
According to the Borel-Weil theorem each linear 
combination of $\Omega_i$ 
\be
\sum_i n_i \Omega^i
\la{young}
\ee
with integer coefficients corresponds to an 
irreducible representation of $SU(N)$. 

When we combine the projection operator (\ref{proj1}) with 
the relation (\ref{m2}) and the Jacobi identity, we find that 
these two-forms $\Omega^i$ can be represented in terms of
the $m_i$,
\be
\Omega^i \ = \ Tr(m_i [ dm_k , dm_k ] ) \ = \ (m_i , dm_k \wedge
dm_k )
\la{kiri}
\ee
Explicitly, in components 
\be
\Omega_{i, \mu \nu} \ = \ f^{abc}m^a_i \partial_\mu m^b_k 
\partial_\nu
m^c_k
\la{kiri2}
\ee

Next, we proceed to consider a generic Lie algebra 
element $v = v^a T^a$. We define the (infinitesimal) 
adjoint action $\delta^i$ of the $m_i$ on $v$ by 
\be
\delta^i v \ = \ [v , m_i ] 
\la{deltai}
\ee
Using (\ref{proj1}) and by summing over $i$ we find that this
yields (up to a sign) a projection operator
to a subspace which is orthogonal to the maximal torus
and is spanned by the $m_i$,
\be
(\delta^i)^2 v \ = -\ v + m_i (m_i , v) 
\la{nilp}
\ee

We shall also need a (local) basis of Lie-algebra 
valued one-forms in the subspace to which (\ref{nilp})
projects. For this we first use (\ref{redu}) to 
conclude that for the symmetric combination
\be
\{ d m_i , m_j \} \ + \ \{ dm_j , m_i \} \ = \ d_{ijk} dm_k
\la{rep1}
\ee
and using (\ref{dijk}) we invert this into
\[
dm_k \ = \ \frac{N}{2(N-2)} d_{kij} ( \ \{ d m_i , m_j \} \ 
+ \ \{ dm_j , m_i \} \ )
\]
Consequently the symmetric combination (\ref{rep1}) yields 
the $SO(N-1)$ vector one-form
\be
X^i \ = \ X^i_\mu dx^\mu \ = \ \partial_\mu m^a_i T^a dx^\mu \ = \
\frac{N}{2(N-2)} d_{ijk} ( \ \{dm_j , m_k \} + \{
dm_k , m_j \} \ )
\la{vec1}
\ee
The antisymmetric combination yields a
$SO(N-1)$ antisymmetric tensor one-form
\ba
Y^{ij} \ & = & \ Y^{ij}_\mu dx^\mu \ = - Y^{ji}_\mu dx^\mu \ 
= \ \{ dm_i , m_j \} \ - 
\ \{ dm_j , m_i \} 
\\
& = & \ d^{abc} (\partial_\mu m^a_i m^b_j 
- \partial_\mu m^a_j m^b_i ) \ T^c dx^\mu  
\la{asym1}
\ea
Finally, we define the $SO(N-1)$ symmetric tensor one-form
\be
Z^{ij} \ = \ Z^{ij}_\mu dx^\mu \ = \ Z^{ji}_\mu dx^\mu \ = \ 
[dm_i , m_j ] \ = \ f^{abc} 
\partial_\mu m^a_i m^b_j \ T^c dx^\mu
\la{sym1}
\ee 
Under $SO(N-1)$ this decomposes into the traceless symmetric 
tensor representation and the trace {\it i.e.} singlet 
representation, but we use it as is.

Observe that that (\ref{vec1}), (\ref{asym1}) and (\ref{sym1})
are the {\it only} invariant one-forms that can be constructed
using the variables $m_i$ and natural $SU(N)$ invariant concepts.
In particular, $X^i$, $Y^{ij}$ and $Z^{ij}$ 
are orthogonal to $m^i$, hence they determine a basis in the 
corresponding subspace $SU(N)/ U(1)^{N-1}$. We can identify them 
as a basis of roots in $SU(N)$.

The dimension of the $SO(N-1)$ vector 
representation (\ref{vec1}) is $N-1$. The dimension 
of the antisymmetric tensor representation (\ref{asym1})
is $\half(N-1)(N-2)$. The sum of these coincides with 
the dimension of (\ref{sym1}),
\[
N-1 \ + \ \half ( N-1 ) (N-2 )
\ = \ \half N(N -1)
\]
Moreover, we find that the $U(1)$ generators (\ref{deltai}) 
map $X^i_\mu$ and $Y^{ij}_\mu$ into $Z^{ij}_\mu$ and 
{\it vice versa}, 
\be
\delta^i X^j \ = \ Z^{ij} 
\la{Xact}
\ee
\be
\delta^i Y^{jk}  \ = \ d^{ikl} Z^{jl} \ - \ d^{ijl} Z^{kl}
\la{Yact}
\ee
and
\[
\delta^i Z^{jk} \ = \ - \frac{1}{N} ( \delta_{ij} 
X^k + \delta_{ik} X^j ) 
\]
\be 
+ \ \frac{1}{4} ( d_{jkl} d_{lin} - d_{jil}
d_{lkn} - d_{kil} d_{ljn}) X^n \ + \ 
\frac{1}{4} ( d_{jkl} Y^{il} + d_{jil} Y^{kl} + d_{kil}
Y^{jl} )
\la{Zact}
\ee
We note that this determines a natural complex structure. 

We have now constructed four different sets of $SU(N)$ 
Lie-algebra valued forms (in $\mu$) from the $m_i$. 
Each of these four sets induces an irreducible representation 
of $SO(N-1)$, they decompose into the vector, the 
antisymmetric tensor, and the traceless symmetric 
tensor plus scalar representations of $SO(N-1)$. 
The cotangent bundle to the
co-adjoint orbit $ SU(N) / U(1)^{N-1} $ is  
spanned by the one-forms (in $\mu$) $X^i , \ Y^{ij}$ and $Z^{ij}$. 

By construction $(m_i , X^i , Y^{ij}, Z^{ij})$ yields a 
complete set of basis states for the $SU(N)$ Lie algebra, and 
can be used to decompose generic $SU(N)$ connections. 
For this we need appropriate dual variables 
that appear as coefficients. We first note that
the connection $A^a_\mu$ is a $SU(N)$ Lie-algebra valued 
one-form, and the $SO(N-1)$ acts on it trivially.  Consequently
the variable which is dual to $m_i$ must be a one-form 
which transforms as a vector under $SO(N-1)$. We call it 
$C^i$. The variables which are dual to the $(X^i , Y^{ij}, 
Z^{ij})$ are zero-forms, and in order to form invariant
combinations they must decompose under the action 
of $SO(N-1)$ in the same manner as $X^i , \ Y^{ij}$ and $Z^{ij}$. 
Since we have also found a natural complex structure which is 
determined by the $\delta^i$, 
this suggests that we denote these dual variables by 
$\phi^{ij} = \rho^{ij} + i \sigma^{ij}$. Here $\rho^{ij}$ 
is dual to the $Z^{ij}$ and can be decomposed into a 
traceless symmetric tensor and a singlet under $SO(N-1)$.
The $\sigma^{ij}$ is dual to the $X^i$ and $Y^{ij}$. 
It decomposes into a vector and an antisymmetric tensor 
under $SO(N-1)$. The ($SO(N-1)$ invariant) $SU(N)$ 
connection $A_\mu^a$ then decomposes into
\be
A \ = \ C \cdot m \ + \ (1 + \rho) [dm ,  m ] \ + 
\ \sigma \{ dm , m \}
\la{finalA}
\ee
{\it Exactly} in four dimensions this contains 
the correct number of independent variables 
for a general $SU(N)$ connection.

The ensuing curvature two form $F = dA + AA$ is obtained  
by a direct substitution. Of particular interest is 
the structure of $F$ in the direction $m_i$,
the maximal torus, 
\be
F^a_{\mu\nu} \ = \ m^a_i( \partial_\mu C^i_\nu - 
\partial_\nu C^i_\mu ) \ - \ m^a_i \Omega^i_{\mu\nu} 
\ + \ ...
\la{F}
\ee
Here the $\Omega^i$ are the Kirillov two forms (\ref{kiri}),
the terms that we have not presented explicitely 
depend on $\phi_{ij}$ and/or are in 
the direction of $SU(N) / U(1)^{N-1}$. (If 
we evaluate the curvature two form for the $SU(N)$ Cho 
connection (\ref{cho}), we find exactly the terms in 
(\ref{F}).) Consequently $F$ is a generating 
functional for the Kirillov two forms. In particular, 
for a flat connection we have 
\[
dC^i = \Omega^i
\]

In \cite{fad1}, using Wilsonian renormalization
group arguments we suggested that for $SU(2)$
(\ref{su2}) the following action
\be
S \ = \ \int d^4x \ (\partial_\mu n)^2 \ + \
\frac{1}{e^2} ( n , dn \wedge d n)^2 
\la{act1}
\ee
may be relevant in the infrared 
limit of Yang-Mills theory.
This is interesting since (\ref{act1}) 
describes knotlike configurations as solitons. The 
(self)linking of these knots is computed by the 
Hopf invariant
\be
Q \ = \ \int F \wedge A
\la{Qh}
\ee
where
\[
F \ = \ dA \ = \ ( n , d n \wedge dn )
\]

The present construction suggests a natural generalization
of (\ref{act1}) to $SU(N)$,
\be
S \ = \ \int d^4x \ (\partial_\mu m_i )^2 \ + \ 
\frac{1}{e_i^2} ( \ [\partial_\mu m_i , 
\partial_\nu m_i ] \ )^2 
\la{act2}
\ee
which reduces to (\ref{act1}) for $N=2$.
Since $\pi_3( SU(N) / U(1)^{N-1} )= Z$ we expect 
that in the general case of $SU(N)$ we also have 
solitons. It would be interesting to 
understand their detailed structure.  

As an example we consider $SU(3)$, the gauge group of strong 
interactions. There are two $SU(3)$ Lie-algebra valued 
vectors $m^a_i$ which we denote by $m^a$ and $n^a$ respectively.
We have $ m^2  =  n^2  = 1$ and $m^a n^a = 0$, and 
(\ref{A1}) becomes
\ba
A^a_\mu \ & = & \ B_\mu m^a + C_\mu n^a + 
f^{abc} \partial_\mu m^b m^c 
+ f^{abc} \partial_\mu n^b n^c 
\la{cho3}
\\
& + & \rho_{mm} f^{abc}\partial_\mu m^b m^c + \rho_{mn} f^{abc}
\partial_\mu m^b n^c + \rho_{nn} f^{abc} \partial_\mu
n^b n^c
\la{f3}
\\
& + & \sigma_{mm} \partial_\mu m^a + \sigma_{nn} \partial_\mu n^a
+ \sigma_{mn} d^{abc} \partial_\mu m^b n^c
\la{d3}
\ea
Here (\ref{cho3}) is the SU(3) Cho connection, and (\ref{f3}),
(\ref{d3}) are the components of $A^a_\mu$ in the direction of 
the orbit $SU(3)/U(1)\times U(1)$. These components transform into 
each other under the action of the operators
\[
\delta_m \  =  \ [ \ \bullet \ , m \ ] \ \ \ \ \ \ \ 
\delta_n \  =  \ [ \ \bullet \ , \ n \ ]
\]
In the Gell-Mann basis the vectors $m$ and $n$ satisfy 
\bastar
[m,n] \ & = & \ 0
\ \ \ \ \ \ \ \ \ \ \ \ \ 
\{ m , m \} \  =  \ \frac{1}{\sqrt{3} } n
\\
\{ n , n \} \ & = & \ - \frac{1}{\sqrt{3} } n
\ \ \ \ \ \ \ 
\{ m , n \} \  =  \ \frac{1}{\sqrt{3} } m
\eastar
Notice that $n$ is represented in terms of $m$, 
a unit vector with six independent field degrees of freedom
to parametrize $SU(3)/U(1) \times U(1)$.

The action of $\delta_m$ and $\delta_n$ on $SU(3)/U(1) \times U(1)$
can be diagonalized. We find that 
$\delta_m$ corresponds to the adjoint action of the Cartan 
element $\lambda_3$ and $\delta_n$ to the adjoint action of the 
Cartan element $\lambda_8$. This decomposes the $SU(3)/U(1) 
\times U(1)$ components of $A^a_\mu$ into $U(1)$ multiplets. 
In particular, $\delta_m^2 + \delta_n^2$ is a projection 
operator onto the basis one-forms of $SU(3)/U(1) \times U(1)$.
Finally, the Kirillov symplectic two-forms are
\[
\Omega_m \ = f^{abc}m^a (\partial_\mu m^b 
\partial_\nu m^c + \partial_\mu n^b \partial_\nu
n^c)
\]
and 
\[
 \Omega_n \ = \ f^{abc}n^a (\partial_\mu m^b 
\partial_\nu m^c + \partial_\mu n^b \partial_\nu
n^c )
\]
and the action (\ref{act2}) is
\[
S \ = \ \int d^4x \ \left( \ (\partial_\mu m )^2  + 
(\partial_\mu n)^2 \ + \ \frac{1}{e_m^2} (f^{abc} \partial_\mu m^b
\partial_\nu m^c)^2 \ + \ \frac{1}{e_n^2} (f^{abc} \partial_\mu n^b
\partial_\nu n^c)^2 \ \right)
\]

\vskip 0.3cm
In conclusion, we have presented a group theoretical
decomposition of four dimensional $SU(N)$ connection
$A^a_\mu$, which we argue is complete in the sense described
in \cite{fad1}. This decomposition involves a number of
natural group theoretical concepts, and in particular relates
the $SU(N)$ curvature two form to the Kirillov symplectic
two forms on the $SU(N)$ coadjoint orbits. Curiously, we also
find that the components of $A$ in the direction of these orbits
can be decomposed according to irreducible representations
of $SO(N-1)$, with a natural complex structure. Our construction
suggests a new class of nonlinear models generalizing the model
first proposed in \cite{fad2}. These models may have
interesting properties, including the possibility of 
solitons with nontrivial topological structures \cite{nature}.

\vskip 0.5cm
We thank W. Kummer, J. Mickelsson and M. Semenov-Tyan-Shansky  
for discussions and comments. We also thank the
the Erwin Schr\"odinger Institute for hospitality.

\end{document}